
\input harvmac

\def\np#1#2#3{Nucl. Phys. B{#1} (#2) #3}
\def\pl#1#2#3{Phys. Lett. {#1}B (#2) #3}

\def\physrev#1#2#3{Phys. Rev. {D#1} (#2) #3}

\def\prep#1#2#3{Phys. Rep. {#1} (#2) #3}

%
%
\ifx\epsfbox\UnDeFiNeD\message{(NO epsf.tex, FIGURES WILL BE IGNORED)}
\def\figin#1{\vskip2in}
\else\message{(FIGURES WILL BE INCLUDED)}\def\figin#1{#1}\fi
\def\ifig#1#2#3{\xdef#1{fig.~\the\figno}
\goodbreak\midinsert\figin{\centerline{#3}}%
\smallskip\centerline{\vbox{\baselineskip12pt
\advance\hsize by -1truein\noindent\footnotefont{\bf Fig.~\the\figno:}
#2}}
\bigskip\endinsert\global\advance\figno by1}

\def\ifigure#1#2#3#4{
\midinsert
\vbox to #4truein{\ifx\figflag\figI
\vfil\centerline{\epsfysize=#4truein\epsfbox{#3}}\fi}
\narrower\narrower\noindent{\footnotefont
{\bf #1:}  #2\par}
\endinsert
}


\def\nf{n_F}
\def\nfb{n_{\bar F}}
\def\bnf{{\bar n}_F}
\def\bnfb{{\bar n}_{\bar F}}

\Title{RU-95-29}
{\vbox{\centerline{The Dual of Supersymmetric SU(2k) with an Antisymmetric}
\centerline{Tensor and Composite Dualities} }}
\smallskip
\centerline{M. Berkooz \footnote{*}{e-mail: berkooz@physics.rutgers.edu}}
\smallskip
\centerline{Department of Physics and Astronomy}
\centerline{Rutgers University}
\centerline{Piscataway, NJ 08855-0849}
\vskip 2cm
\baselineskip 18pt

\noindent

We suggest a dual to an $SU(2k)$ Susy gauge theory containing an
 antisymmetric tensor, $\nf$ fundamentals and $\nfb$ anti-fundamentals.
 This is done by expanding the theory into an equivalent description with
 two gauge groups and then performing known duality tranformations on each
 gauge group separately. Chiral operators, mass perturbations and flat
 directions are discussed.

\Date{5/95}
\nref\sem{N. Seiberg, hep-th/9411149 , RU-94-82, IASSNS-HEP-94/98}%
\nref\ads{I. Affleck, M. Dine and N. Seiberg, \np{241}{1984}{493};
 \np{256}{1985}{557}}%

\nref\nonren{N. Seiberg, hep-ph/9309335, \pl{318}{1993}{469}}%
\nref\nati{N. Seiberg, hep-th/9402044, \physrev{49}{1994}{6857}}%
\nref\ils{K. Intriligator, R.G. Leigh and N. Seiberg, hep-th/9403198,
\physrev{50}{1994}{1092}; K. Intriligator, hep-th/9407106,
\pl{336}{1994}{409}}%
\nref\swi{N. Seiberg and E. Witten, hep-th/9407087, \np{426}{1994}{19}}%
\nref\swii{N. Seiberg and E. Witten, hep-th/9408099,
\np{431}{1994}{484}}%
\nref\intse{K. Intriligator and N. Seiberg, hep-th/9408155,
\np{431}{1994}{551}}%
\nref\kutasov{D. Kutasov, hep-th/9503086, EFI-95-11}%
\nref\tau2{O. Aharony, J. Sonnenschein and S. Yankielowicz, hep-th/9502013,
 TAUP-2232-95}%
\nref\kutsch{D. Kutasov and A. Schwimmer, EFI-95-11, WIS/4/95,
hep-th/9505004}%
\nref\rlms{R. Leigh and M. Strassler, RU-95-2, hep-th/9503121}%
\nref\kenphil{K. Intriligator and P. Pouliot, RU-95-23, hep-th/9505006}%
\nref\so{K. Intriligator and N. Seiberg, RU-95-3, IASSNS-HEP-95/5,
hep-th/9503179}%
\nref\power{N. Seiberg, The Power of Holomorphy -- Exact Results in 4D
SUSY Field Theories.  To appear in the Proc. of PASCOS 94.
hep-th/9408013, RU-94-64, IASSNS-HEP-94/57}%
\nref\nfour{{H. Osborn, \pl{83}{1979}{321};A. Sen, hep-th/9402032,
\pl{329}{1994}{217}; C. Vafa and E. Witten, hep-th/9408074,
\np{432}{1994}{3}}}%

\nref\banks{T. Banks and A. Zaks, \np{196}{1982}{189}}%

\lref\svgc{M.A. Shifman and A. I. Vainshtein, \np{296}{1988}{445};
A. Yu. Morozov, M.A. Olshanetsky and M.A. Shifman, \np{304}{1988}{291}}%
\nref\nsvz{V.A. Novikov, M.A. Shifman, A. I.  Vainshtein and V. I. Zakharov,
\np{223}{1983}{445}; \np{260}{1985}{157}; \np{229}{1983}{381}}%
\nref\cern{D. Amati, K. Konishi, Y. Meurice, G.C. Rossi and G. Veneziano,
\prep{162}{1988}{169} and references therein}%
\lref\sv{M.A. Shifman and A. I. Vainshtein, \np{277}{1986}{456};
\np{359}{1991}{571}}%

\lref\om{C. Montonen and D. Olive, \pl {72}{1977}{117}}

\vfill
\eject

\newsec{Introduction}

As has been recently demonstrated, the holomorphic structure of Susy gauge
theories allows one to calculate some of their I.R. properties exactly
\sem \refs{\nonren-\nfour} (for earlier work see \ads\refs{\nsvz-\sv}).
In particular some of them can be given an equivalent low energy description
that interchanges weak and strong coupling regions, generalizing the
Olive-Montonen electric-magnetic duality \om. Most existing $N=1$ examples
of this duality have a common flavor to them in the sense that the dual model
is similar to the original model with a different number of colors and
additional gauge invariant fields. There is, however, no necessity that this
will be the general case. Accepting the notion that non-abelian gauge degrees
of freedom are almost fictitious in the infra-red, one may have a dual,
or equivalent description, with an arbitrary gauge groups structure.

We will suggest that an $SU(2k)$ Susy gauge theory with an antisymmetric
tensor and appropriate number of quarks and anti-quarks has an equivalent
description in terms of a theory with two gauge groups and therefore a chain
of duals with two gauge groups. By construction, anomaly matching is immediate
and so is partial identification of the chiral ring. We also discuss mass
perturbations and flat directions.

\newsec{The Electric Theory}

\medskip

The theory in question is an $SU(N=2k)$ gauge theory with $\nf$ quarks ($Q$)
in the fundamental rep., $\nfb$ antiquarks ($\bar Q$) in the anti-fundamental
rep., and an antisymmetric tensor ($A$) \ads. Anomaly cancelation requires
$N=\nfb-\nf+4$.

There are two ordinary U(1) global symmetries and one ${U(1)}_R$. We will be
interested, however, in the theory with a superpotential $W=Pf(A)$. Let us
assume that we do not add this superpotential and discuss the status of the
operators $Pf(A)QA^{-1}Q$. By the following pragmatic and very porous
argument, we have to set these operators to zero.

There are 3 options:

1. $Pf(A)QA^{-1}Q$ is elementary in the dual theory. Let us look at the
limit $\nf,\nfb\rightarrow\infty$ at a fixed ratio. Some $\sum U(1)$ anomaly
term will scale like $N^3$ and $\sum {U(1)}^3$ will scale like $N^5$. It is
unclear how to cancel these contributions. We are justified in taking this
limit since the equations for the anomaly matching are algebraic. If the
anomalies match in a limited region of $\nf$ and $\nfb$ then they can be
analytically continued to match in non-physical regions. We can thus take
the limit $\nf,\ \nfb\rightarrow\infty$ at fixed ratio for arbitrary such
ratio.

2. It is a composite object. In this case we expect all of its ${U(1)}$
charges to scale like $N$ in the dual theory, whereas they do not do so
in the original theory\footnote{$^1$}{In any case, in appendix 1 we
construct the dual without assuming any $W$ and derive it}.

3. It does not exist. This is exactly what $W=Pf(A)$ does.

\medskip
Adding this superpotential, we are left with a theory that has the
following $U(1)$ charges:

$$\vbox{\settabs 4 \columns
\+          &${U(1)}_1$         &${U(1)}_R$                             \cr
\+$Q$       &$1$                &$1+{8-6N-N^2\over N(\nf+\nfb)}$   \cr
\+$\bar Q$  &$-{\nf\over\nfb}$  &$1+{8-6N-N^2\over N(\nf+\nfb)}$   \cr
\+$A$       &$0$                &$4/N$                             \cr
}$$

Asymptotic freedom is lost at $6N-\nf-\nfb-(N-2)=0$. For reasons that will
be clear later, we will be interested primarily in the region
$3\nf-12>\nfb>2\nf-1$ (in this region the theory is asymptotically free).
Note also that in this range, at generic points along the flat directions,
the gauge group is completely Higgsed.

There is no additional dynamically generated superpotential when the gauge
group is completely Higgsed. In this case a dynamically generated
superpotential should be understood as an instanton effect and we can
expand the superpotential in instanton powers. Imposing the various
symmetries then forbids any additional superpotential \refs{\nonren-\ils}.

As in \sem, one can make the following arguments in favor of the existence
of a non-abelian Coulomb phase:

1. In the limit where $\nf$ and $\nfb$ very large, and $\nf+\nfb$ very close
but smaller then $5N+2$ (this region of $\nf$ and $\nfb$ is excluded from
the region above, though) then the 2 loop term in the $\beta$ function can
balance the 1 loop $\beta$ function, yielding a non trivial, interacting,
field theory in a perturbative regime of the coupling constant \banks.

2. The theory has classical flat directions in which $A=0$ and so are
$\nfb-\nf$ of the antiquarks. The D-term equations are then the usual ones
for $SU(N)$ with $\nf$ flavors. We can break $SU(N)$ to $SU(k)$ and be left
with $\nf-N+k+(N-k)=\nf$ quarks ($A$ also contributes fundamentals of
$SU(k)$), $\nfb-N+k=\nf+k-4$ antiquarks and an antisymmetric tensor. The
theory is not asymptotically free when $4k+6-2\nf\leq0$, i.e.,
$k_{max}\leq{\nf-3\over2}$. We would expect, at least, an $SU(k_{max})$
subgroup to be in a non-abelian Coulomb phase at the origin.

\newsec{The Expanded Model}

One can go and directly find other models satisfying the 't Hooft anomaly
matching conditions and check if they are dual to this
model\footnote{$^2$}{Appendix 1 contains a series of arguments that
determines the dual directly}, or one can use the following shortcut.

As described in \sem, in SQCD, when $N_f=N_c+1$, mesons and baryons saturate
the anomaly matching conditions, and constitute a valid description of the
low energy physics. Such a point exists also for
$Sp(N_2)$\footnote{$^3$}{The notation is such that $N_2$ is the dimension
of the fundamental. The dimension of the group is then
${N_2(N_2+1)\over2}$} \kenphil. Indeed, the mesons alone saturate the
anomaly, which is fortunate since the baryons decompose into an
antisymmetrized product of mesons. We would therefore like to suggest
that the model at hand is equivalent to the following model:

$$\vbox{\settabs 4 \columns
\+                &$SU(N)$             &$Sp(N_2,U)$             \cr
\+$\hat Q$        &$N$                 &$1$                &$\nf$          \cr
\+$\hat{\bar Q}$  &$\bar N$            &$1$                &$\nfb$         \cr
\+$X$             &$N$                 &$N_2$                              \cr
}$$
where $N=\nfb-\nf+4$, $N_2=\nfb-\nf$ and $A\sim XJX$ where $J$ is the
symplectic form.

The equivalence between the two models is the following. We first solve
the D-term constraints of $Sp$ \kenphil. Choose a maximal lagrangian
space\footnote{$^4$}{A subspace of $C^{N_2}$ on which the symplectic form
vanishes} in the space spanned by the $Sp$ vectors in $X$, and diagonalize
that part of $X$ using $SU(N)$ and $Sp(N_2)$ transformations. The D-term
constraints then imply that we have the same diagonal values in the conjugate
lagrangian subspace (up to an allowed $SU(N)$ transformation). This defines
a $1-1$ correspondence between $A$, constrained by $W=Pf(A)$, and $X^2$,
with $X$ transforming as above.

This, however, is not entirely correct. This correspondence, appended with
$\hat Q=Q$ and ${\bar Q}=\bar Q$, does not map $D^{SU}$ in the original
theory to $D^{SU}$ in the expanded theory. This can be fixed by twisting the
transformation by $A=GX^2G^T,\ Q=G\hat Q,\ \bar Q=G^{-1}\hat{\bar Q}$ where
$G\in SL(N,C)$, depends on $Q,\bar Q, A$, and acts on the $SU(N)$ indices of
the fields. To find $G$ we require that it intertwines the $D$-terms of the
$SU(N)$ in the two theories.

More precisely: The D-term for $SU(N)$ in the original theory is
$$-{\bar Q}^*{\bar Q}^T+QQ^\dagger +2AA^\dagger=c_1 I$$ and for the expanded
theory it is $$-{\hat{\bar Q}}^*{\hat{\bar Q}}^T+\hat Q{\hat Q}^\dagger +
XX^\dagger = c_2 I$$ and, after eliminating $Q$ and $\hat Q$, the equation
obtained for the twist matrix is
$$-{G^{-1}}^{*}{\hat{\bar Q}}^*{\hat{\bar Q}}^TG+
G{\hat{\bar Q}}^*{\hat{\bar Q}}^TG^\dagger-GXX^\dagger G+
2GX^2G^TG^*{X^2}^\dagger G^\dagger=c_1 I + c_2 GG^\dagger$$

There are $N^2$ real equations (the equation is automatically hermitian) and
$N^2$ real unknown variables (the numbers of parameters in $G$ is
$dim(SL(N,C)/SU(N))$ and $c_1$). We have not been able to solve the relation,
but typically we expect that there is a 1-1 map from one set of elementary
quarks to another, which, by construction, does not alter the values of
composite objects. Written in terms of gauge invariant objects this is a
$1-1$ holomorphic map from one moduli space of vacua to the other,
intertwining all constraints between gauge invariant holomorphic objects.

The physics here is that of confinement \kenphil\ and the two theories are
equivalent in the I.R., only after the massive bound states in the $Sp$
sector have decoupled from the theory. The gauge dynamics in $Sp$ is
strictly strong; the fixed point we flow to is at finite value of the $SU$
coupling constant but at infinite value of the $Sp$ coupling constant.

We suggest that a similar construction can serve as a general guideline for
constructing sets of duals. For example, one could think of replacing a
symmetric tensor of $SU(N)$ with an $SO(N+4)$ gauge group \so\ and a new
fundamental field transforming as $N\times(N+4)$ under $SU(N)\times SO(N+4)$.
Indeed at $N'_c=N'_f+4$, $SO(N'_c)$ has several branches of vacua. In one
type of these branches there is no superpotential (and on the other there
is no ground state) and the symmetric meson of this theory is the correct
description in the I.R. An adjoint in $SU(N)$ may require expanding with
other $SU(N')$ gauge groups \refs{\kutasov-\kutsch}.

\newsec{The Dual Theories}

At this stage one has a model which contains only (anti)fundamentals in each
gauge group. We can then obtain suggestions for a dual model by first
dualizing the $SU$ sector by the known $SU$ duality \sem\ and then dualizing
the $Sp$ sector by a similarly suggested duality \sem\kenphil; clearly this
satisfies the anomaly matching conditions. It also provides the translation
dictionary between bound states in the original, expanded and dual theories;
given a bound state in the original theory, translate it into the expanded
theory and then construct, in stages, its counterparts. When dualizing one
gauge group, some of the mesons generated are charged under the second gauge
group but this does not affect the translation table.

After dualizing $SU(N)$ \sem\ one obtains:
$$\vbox{\settabs 4 \columns
\+      &$SU(N_3=\nf-4)$     &$Sp(\nfb-\nf,U)$   \cr
\+$q_1$ &$N_3$               &$1$           &$\bnf$                \cr
\+$q_2$ &${\bar N}_3$        &$1$           &$\bnfb$               \cr
\+$q_4$ &$N_3$               &$N_2$                                \cr
\+$q_5$ &$1$                 &$N_2$         &$\nfb$                \cr
\+$M$   &$1$                 &$1$           &$\nf\times\nfb$       \cr
}$$
with a superpotential $q_2q_5q_4+q_1q_2M$. The notation $\bnfb$ ($\bnf$)
denotes an anti-fundamental under the flavor group $SU(\nfb)$ ($SU(\nf)$).

Note that the $Sp$ group is not driven to the range in which it generates a
superpotential \kenphil. Even if $q_2$ has maximal rank, it can mass up
$\nf-4$ of the $Sp$ quarks at most, leaving $\nfb-\nf+4$ of them massless.
It is remarkable, however, that by dualizing $SU$ we have already transformed
$Sp$ from a region with strong coupling dynamics in the flat directions to a
region without. We will have more to say on this issue in the following
section, but for now we notice that the choice of the range of $\nf,\nfb$
that was done before ensures that both gauge groups are asymptotically free.

One can dualize $Sp$ \sem\kenphil \space  in this model and, after
integrating out some degrees of freedom \refs{\nati-\ils}, obtain:

$$\vbox{\settabs 4 \columns
\+      &$SU(\nf-4)$           &$Sp(N_4=2\nf-8)$   \cr
\+$p_1$ &$N_3$                 &$1$           &$\bnf$                  \cr
\+$p_2$ &$1$                   &$N_4$         &$\bnfb$                 \cr
\+$p_3$ &${(N_3\times N_3)}_A$ &$1$                                    \cr
\+$p_4$ &${\bar N}_3$          &$N_4$                                  \cr
\+$N$   &$1$                   &$1$           &${(\nfb\times\nfb)}_A$  \cr
\+$M$   &$1$                   &$1$
          &$\nf\times\nfb$         \cr
}$$
with a superpotential $Np_2p_2+Mp_1p_4p_2+p_3p_4p_4$.

The last term, which is automatically generated, eliminates the meson
$p_1p^2_4p_1$ from the chiral ring.

Note that the regions in which the expanded and two dual theories are
asymptotically free are different (but overlapping) regions of $\nf$ and
$\nfb$. One could argue that the descriptions are dual on the regions of
overlap (which is the region we have been restricting ourselves to so far)
but there is another possible solution to this puzzle, which we will present
in the next section.

\newsec{Chiral Operators}

As noted above, identifications of chiral operators are almost automatic.
We will demonstrate the translation for a few operators:

$$\vbox{\settabs 3 \columns
\+Original theory     &Expanded theory          &Dual 1  	\cr
\+$Q\bar Q$           &$Q\bar Q$                &$M$             \cr
\+$\bar Q A\bar Q$    &$\bar QX^2\bar Q$        &$q_5^2$         \cr
\+${\bar Q}^N$       &${\bar Q}^N$             &${q_2}^{\nf-4}$ \cr
\+$Q^kA^{N-k\over2}$ &$Q^k{(XJX)}^{N-k\over2}$
&$q_1^{\nf-k}{(q_4Jq_4)}^{k-4\over2}$\cr
}$$

Translating into Dual 2 is also simple; replace any $q_5^2$ by $N$ and
$q_4^2$ by $p_3$ (noting also that the corresponding dual mesons of the
$Sp$ sector, $p_2^2$ and $p_2p_4$, are redundant). Since $q_2$ has been
integrated from the theory via its equations of motion, which are
$q_2=p_2p_4$, the baryon ${\bar Q}^N$ now corresponds to
${(p_2p_4)}^{\nf-4}$.

The $U(1)$ charges of these operators are:
$$\vbox{\settabs 3 \columns
\+Original theory    &${U(1)}_1$          &${U(1)_R}$	\cr
\+$Q\bar Q$          &$1-{\nf\over\nfb}$  &$2+2{8-6N-N^2\over N(\nf+\nfb)}$\cr
\+$\bar Q A\bar Q$   &$-2{\nf\over\nfb}$
&$2+2{8-6N-N^2\over N(\nf+\nfb)}+{4\over N}$   \cr
\+${\bar Q}^N$       &$-{\nf\over\nfb}N$  &$N+{8-6N-N^2\over\nf+\nfb}$     \cr
\+$Q^kA^{N-k\over2}$ &$k$
  &$k(1+{8-6N-N^2\over\nf+\nfb})+{4(N-k)\over2N}$ \cr
}$$

It is simplest to examine these relations in the limit that $N,\nf,\nfb$ are
taken to infinity at fixed ratios. In this case the charges are

$$\vbox{\settabs 3 \columns
\+Original theory    &${U(1)}_1$          &${U(1)}_R$                \cr
\+$Q\bar Q$          &$1-{\nf\over\nfb}$  &$2-2{N\over \nf+\nfb}$     \cr
\+$\bar Q A\bar Q$   &$-2{\nf\over\nfb}$  &$2-2{N\over \nf+\nfb}$     \cr
\+${\bar Q}^N$       &$-{\nf\over\nfb}N$  &$N(1-{N\over\nf+\nfb})$    \cr
\+$Q^kA^{N-k\over2}$ &$k$
&$k(1-{N\over\nf+\nfb})+{4(N-k)\over2N}$ \cr
}$$
and one observes that there is a continuum of R symmetries that are allowed
as the R symmetry of the conformal field theory (CFT) to which the theory
flows in the infrared, i.e., these choices of charges that do not violate
any unitarity bound (the only restriction is $2-2{N\over{\nf+\nfb}}>{2\over3}$
 which is automatic in the region of $\nf$, $\nfb$ we are looking at). More
specifically, we can define ${U(1)}_R^{new}={U(1)}_R+c{U(1)}_1$ where $c$
is a real number, bounded from above and below by the unitarity bound \sem.
To remind the reader, the unitarity bound is $D(\phi)={3\over2}R(\phi)$
where $D(\phi)$ is the scaling dimension of any gauge invariant chiral
field $\phi$; $D>1$ for interacting spinless gauge invariant chiral fields
and $D=1$ for non interacting.

This is unlike the situation in most existing examples. For example in
$SU(N'_c)$ with $N'_f$ flavors \sem, a discrete charge conjugation makes it
natural to assign equal R charges to quarks and antiquarks, pinpointing the
R symmetry with which to calculate the scaling dimensions.

Note, however, that none of the allowed R symmetries here can be identified
with the R symmetry in the SQCD case.

We can now also suggest another solution to the problem presented in the
section before. The unitarity bound, for large $\nf$ and $\nfb$, is
$\nfb<5\nf$ (to check non leading terms we have to include the 2nd loop term
in the $\beta$ function, this contains the scaling dimensions of non gauge
invariant operators, which we don't know). This is also the limit in which
dual 2 looses asymptotic freedom. So one is led to conclude that the
description in terms of dual 2 is valid up to that point. This is similar to
what happens in other existing examples. Dual 1, however, behaves differently.
The $SU$ gauge group is dual 1 apparently looses asymptotic freedon earlier,
at $\nfb=3\nf$, however, no unitarity bound is violated at this transition.
One is then led to speculate that it still flows to a CFT, at a regime in
which perturbation theory is meaningless, the usual calculation of asymptotic
freedom is not reliable, and the theory is not free in the I.R.

\newsec{Flat Directions}

We will show that, in the flat directions of the expanded model and dual 1,
the submanifolds of generic points are the same. Let us for a minute neglect
the D-terms of the $Sp$ group. Then we just have an $SU$ theory and its dual,
and the equivalence of their flat directions has been established in \sem.
In our notation we have two manifolds, on one of them, ${\cal M}$, there is
a set of coordinates $\{Q^kX^{N-k},Q\bar Q,X\bar Q,{\bar Q}^N\}$ satisfying
some set of holomorphic constraints and on the other one there is a set of
coordinates $\{q_1^{\nf-k}q_4^{k+4},M,q_5,q_2^{\nf-4}\}$ and, by $SU(N)$
duality, these are the same coordinates on the same manifold. The $Sp$
symmetry acts on this manifold, and again by construction, $Sp^c$
(complexified $Sp$) acts in the same way on the two manifolds. It is now
easy to see that the moduli spaces of our theories, with gauged $Sp$, are
isomorphic at generic points since the generic points of the moduli spaces
are both the generic points of ${\cal M}/{Sp(N_2,U)}^c$.

The explicit computation is straightforward. We first argue that the
arguments in \sem\space that lead to the equivalence between $SU$ and its
dual still hold. One may worry that the superpotential used in \sem\space is
not gauge invariant under $Sp$ and is not allowed in our theory. Indeed, the
form presented there, althought reflecting the correct physical picture, is
not $SU(N_F)$ invariant. This is, however, a technical point and one can
write it in a flavor invariant way\footnote{$^5$}{We keep the massive quarks
in order to make the flavor symmetry manifest and then integrate them out.}.

The next step, i.e. that of modding out by $Sp^c$, is nothing but a global
version of the super-Higgs mechanism. Pick up some $Q,\bar Q, X$ in the
extended theory that are allowed by the $D^{SU}$. The orbits of $Sp^c$ are
given by $X\mapsto GX$ where $G\in Sp^c$. We would like to show that in a
generic orbit of $Sp^c$ there is one orbit of $Sp$ that satisfies $D^{Sp}$.
By looking at $Sp$ gauge invariant objects (or by the Higgs mechanism),
this will then give the desired result that the generic points of the moduli
space of vacua are the generic points of ${\cal M}/{Sp^c}$.

We would like to ask, therefore, whether there is a $G\in Sp^c, G\not\in Sp$
such that both $XX^\dagger$ and $GXX^\dagger G^\dagger$ satisfy $D^{Sp}$. By
choosing specific gauges of $Sp$, the question is whether there is such a $G$
such that
$$\left(\matrix{\alpha&0\cr 0&\alpha}\right)=G\left(\matrix{\beta&0 \cr
0&\beta\cr}\right)G^\dagger$$ where $\alpha$ is a real non-degenerate
diagonal matrix of size ${N_3\over2}\times{N_3\over2}$ (non degeneracy is
the requirement that the point is generic).
Direct algebraic manipulations then show that $G$ in fact has to be in
$Sp(N_2,U)$.

The same argument holds for dual 1, concluding the proof.

\newsec{Adding mass perturbations}

We add to the original theory a perturbation $mQ_1\bar Q_1$ and flow to a
theory with one $n^{new}_F=\nf-1,\ n^{new}_{\bar F}=\nfb-1$. In the dual
theories we add a perturbation $mM_{1,1}$. We would like to show that the
dual theory flows to the dual theory with $n^{new}_F,\ n^{new}_{\bar F}$.

For the first dual the computation is identical to one studied by Seiberg
in \sem \space for $SU(N)$ theories.

In the second dual the computation is the following. We add a perturbation
$M_{1,1}$, choose the most symmetric combination and expand about it.
The configuration we choose is

$$p_{1,1}=\left(\matrix{1\cr0\cr .\cr .\cr}\right),\ p_{2,1}=
\left(\matrix{1\cr\vec0\cr0\cr\vec0\cr}\right),\ p_{4,(N_4\times N_3)}=
\left(\matrix{0&0&\ldots&0\cr\vec0&0&\ldots&0\cr1&0&\ldots&0\cr
\vec0&0&\ldots&0\cr}\right)$$
The complexification of $sp(2n,U)$ is
$$\left(\matrix{A&B\cr C&-A^T\cr}\right)$$ $B^T=B,\ C^T=C$.

We wish to discuss its action on pair of vectors
$$\left({\left(\matrix{1\cr\vec 0\cr 0\cr\vec 0}\right),
\left(\matrix{0\cr\vec 0\cr 1\cr\vec 0}\right)}\right)$$ where we have
divided $C^{N_4}$ into 2 lagrangian subspaces of dimension $N_4/2$ (two
upper components vs. two lower components) and in each subspace we have
separated the first coordinate from the rest.

The relevant span is $$\left(\matrix{A&B\cr C&-A^T\cr}\right)
\left(\matrix{1\cr\vec 0\cr 0\cr\vec 0\cr}\right)=
\left(\matrix{a\cr c\cr}\right)$$
$$\left(\matrix{A&B\cr C&-A^T\cr}\right)\left(\matrix{0\cr\vec 0\cr 1\cr
\vec 0\cr}\right)=\left(\matrix{b\cr -\hat a\cr}\right)$$
where $a,\hat a,c,c$ are arbitrary vectors in $C^{N_4\over2}$ with the
single constaint $a_1={\hat a}_1$.

One sees that the only direction that does not receive mass in the Higgs
mechanism in $Sp$ is the direction $$\lambda\left({\left(\matrix{1\cr\vec 0
\cr 0\cr\vec 0}\right),\left(\matrix{0\cr\vec 0\cr 1\cr\vec 0}\right)}\right),
\ \lambda\in C$$
This direction is massed up together with $M_{11}$.

After taking into account the directions that are Higgsed by $SU$ breaking we
are left with gauge groups that are broken to
$SU(N_4)\rightarrow SU(N_4-1),\ Sp(N_3)\rightarrow Sp(N_3-2)$.

$$p_{1,(N_4\times\nf-1)}=\left(\matrix{\leftarrow&v&\rightarrow\cr\cr
&p'_1\cr\cr}\right),\ p_{2,(N_3\times\nfb-1)}=\left(\matrix{\leftarrow&w&
\rightarrow\cr &p'_2&\cr \leftarrow&t&\rightarrow\cr&p'_2}\right)$$

$$p_{4,(N_4\times N_3)}=\left(\matrix{h&\leftarrow&s&\rightarrow\cr i&&p'_4
\cr g\cr1&\leftarrow&higgs&\rightarrow\cr g&&p'_4\cr s\cr}\right),\ p_3=
\left(\matrix{0&\leftarrow&z
&\rightarrow\cr \uparrow\cr -z&&p'_3\cr \downarrow\cr}\right)$$

$$M_{(\nf\times\nfb)}=\left(\matrix{M_{1,1}&\leftarrow&c&\rightarrow\cr
\uparrow\cr e&&M'\cr\downarrow\cr}\right),\ N_{(\nfb\times\nfb)}=
\left(\matrix{0&\leftarrow&b&\rightarrow\cr \uparrow\cr -b&&N'\cr
\downarrow\cr }\right)$$

Examining the superpotential, $Np_2p_2$ spawns the mass term $tb$,
$Mp_1p_2p_4$ spawns $ev$ and $vs$ (and the mass term for $M_{1,1}$) and
$p_3p_4p_4$ spawns $sz$.

\newsec{Discussion}

We have suggested a dual to an $SU(2k)$ Susy gauge theory containing an
antisymmetric tensor, quarks and antiquarks. This model contains some new
features:

1. The dual contains 2 gauge groups and does not look similar to the original
model.

2. There are no discrete symmetries that fix a specific $R$ symmetry as the
one determining the scaling dimenstions.

3. The duality transformation can be understood in terms of other, simpler,
duality transformations. One may therefore expect to have a set of
``primitive dualities'' that can be composed together after expanding the
models and rewriting complicated representations as confined objects built
out of simpler elementary fields. This procedure of expansion of a gauge
theory with complicated representations into one with more gauge groups, but
all matter fields in the fundamental, may be of more general utility.

We hope this example will be useful in gaining a deeper understanding of
duality.

\medskip

\centerline{{\bf Acknowledgments}}

It is a pleasure to thank T. Banks, M. Dine, K. Intriligator and P. Pouliot
for illuminating discussions and comments, and to thank SLAC and the Weizmann
Institute Of Science for their hospitality while completing this work.

\newsec{Appendix 1}

One can construct the 2nd dual directly.
As claimed above we will not take the baryon $Pf(A)Q A^{-1}Q$ to be an
elementary excitation in the dual theory so we are left with
${Q\bar Q}_{\nf\times\nfb}$ and ${\bar QA\bar Q}_{{(\nfb\times\nfb)}_A}$
as natural candidates.

Let us make the following assumptions:

1. All the dual quarks transform as anti-fundamentals under the flavor groups.
We then know that the dual quarks are in different gauge groups and the size
of the groups.

2. The superpotential is of the form $W=\sum M^D_iM_i+W_2$ where $M^D$ ($M$)
are mesons in the dual (original) theory and $W_2$ is independent of $M$
and of the dual quarks. We are not assuming a potential $Pf(A)$ in the
original theory, and we will derive it.

$M^D_i$ are all the mesons in the dual theory (which are not zero by other
equations of motion). These must be in 1-1 correspondence to the mesons,
$M_i$, taken from the original theory (which are all the mesons in the
original theory not set to zero by the equations of motion there). We cannot
have more $M_i^D$ since then we would have mesons in the dual theory that
are not redundant but are not in the electric theory, and we can not have
more $M_i$ since then we will have a non-interacting (electric) meson the
dual theory whereas we can not identify such a non-interacting meson in the
original theory. We are assuming here that a field cannot be made
interacting only through the K\"ahler potential (and we will also
assume that the K\"ahler potential does not break symmetries apparent in
the superpotential).

The assumption on $W_2$ is a simplifying assumption. We can actually make do
with a weaker assumption that the equations of motion originating from $W_2$
do not set to zero any dual mesons, and that therefore the only way of
eliminating them from the chiral ring is by coupling them to mesons from
the original theory. This is the case in existing examples of duality.

By now we know the sizes of the groups and also have information on their
type. Since we don't have any electric meson in ${(\nf\times\nf)}_A$ we must
forbid dual quarks, $q_D$, transforming as $\bnf$ to bind into a gauge
invariant mesons $q_Dq_D$, so $q_D$ must transform under an $SU$ group
gauge group. Since there is no way to cancel the anomaly in the gauge group
containing the dual quarks transforming as $\bnfb$ (without adding more
gauge groups) it must be an $Sp$ group. (We argue that there is also one
dual quark that transforms under $SU(\nfb)$. Suppose there are two such dual
quarks. If there is no dual meson built from these then we have an
$SU(\nfb)\times SU(\nfb)$ symmetry. If we have two such quark then the dual
meson built from them has no particular symmetry properties. It must couple
to an original $\nfb\times\nfb$ meson that contains both the symmetric and
the antisymmetric part and again we have an $SU(\nfb)\times SU(\nfb)$
symmetry in the superpotential.)

So far we have the following fields in the suggested dual

$$\vbox{\settabs 4 \columns
\+      &$SU(N_3=\nf-4)$  &$Sp(N_4=2\nf-8)$           	\cr
\+$p_1$ &$N_3$        &$1$           &$\bnf$    \cr
\+$p_2$ &$1$          &$N_2$         &$\nf$     \cr
\+$M$   &$1$          &$1$           &$\nf\times\nfb$   \cr
\+$N$   &$1$          &$1$           &${(\nfb\times\nfb)}_A$  \cr
}$$

Denote $l_1=number\ of\ fields\ N_3\times N_3-
number\ of\ fields{\bar N}_3\times{\bar N}_3,\ l_2=k^{+}-k^{-}$, where in
$k^{\pm}$ counts the number of fields that transform under
$SU(N_3)\times Sp(N_4)$ such that a $+$($-$) is an (anti)fundamental under
$SU(N_3)$. We will assume that $l_1$ and $l_2$ are fixed as we vary $\nf$ and
$\nfb$. In principal the change can be of the form
$\lfloor f(\nf,\nfb) \rfloor$ but then the equations for the anomaly matching
will not be algebraic and it is not clear how to deal with them.

Define the following U(1) symmetry: In $SU(N_3)$, to every field, except
$p_1$, give charge +1 if it has an $N_3$ index and -1 if it has an $\bar N_3$
index. Cancel the anomalies with $p_1$ and $p_2$. $N$ and $M$ have charges
$-2p_2$ and $1-p_1-p_2$ (charges are denoted with $p_1$ etc.).

The $U(1)$ above is some $\alpha{U(1)}_1+\beta{U(1)}_2$ (where the charges
under ${U(1)}_2$ in the electric theory are $A_{(1)}$,
$Q,{\bar Q}_{({2-N\over(\nf+\nfb})}$) in the electric theory. Then the
charges of mesons imply
$$-{2\nf\over\nfb}\alpha+(1-2{N-2\over\nf+\nfb})\beta=-2p_2$$
$$(1-{\nf\over\nfb})\alpha-2{N-2\over\nf+\nfb}\beta=1-p_1-p_2$$
from which one obtains $$\beta=-{2\over N}(\nf+{l_2N_3}+2l_1N_3+l_2N_4)$$ in
the limit that $\nf,\nfb\rightarrow\infty$ at fixed ratio, which we will
assume from now on.

Already we can see that we will run into problems with this $\beta$.
Currently it is of the form ${a\nf+b\nfb\over N}$. We have to cancel the $N$
in the denominator, otherwise there will be a $1\over N$ pole in
$\sum {U(1)}^3$ electric, but there is no such pole in the magnetic theory
(because these are algebraic equations we can analytically continue them to
a non-physical regime, in this case $N=0$). After eliminating the pole in
$N$ (which fixes $b=-a$) we are left with a $1\over \nf+\nfb$ pole in
$\sum {U(1)}^3$ electric that does not exist in the magnetic. Thus $\beta=0$.
So we have identified ${U(1)}_1$.

Using the above equations with $\beta=0$, and the equations for $\sum U(1)$
as equations for $\alpha$, $l_1$, $l_2$ (where $\alpha$ may depend on $\nf$
and $\nfb$ but $l_1$ and $l_2$ may not) one obtains $l_1=1$, $l_2=-1$. Since
we don't want any $\bnf\times\bnf$ meson, we have no fields of the form
$\bar N_3\times \bar N_3$, fixing $p_3$ in the 2nd dual. We have no
compelling argument why the solution to the constraint $l_2=-1$ is just with
one $p_4$,
it is certainly the simplest solution. We have tried other ways of solving
the constraint (one has the freedom to add adjoints also) and typically one
has magnetic mesons that are not in the electric theory or too many $U(1)$
in the magnetic theory (long before checking flat directions). After having
the solution one can argue the following. Since we have matched the chiral
ring with just one $p_4$, all gauge invariant objects we build with whatever
additional field have to be zero by the equations of motion. Very likely
this means that we can consistently set all the additional fields to zero
and forget about them.

\listrefs
\end